# Effect of the four-sheet Fermi surface on magnetoresistivity of $MgB_2$

I. Pallecchi[1], M. Monni[1,2], C. Ferdeghini[1], V. Ferrando[1], M. Putti[1], C. Tarantini[1], E. Galleani D'Agliano[1,2]

[1] LAMIA-CNR- INFM and Università di Genova, via Dodecaneso 33, 16146 Genova Italy

[2] CNISM, Unità di Genova, via Dodecaneso 33, 16146 Genova Italy

PACS: 74.70.Ad, 75.47.Np

**Abstract**

Recent experimental data of anisotropic magnetoresistivity measured in $MgB_2$ films have shown an intriguing behaviour: the angular dependence of magnetoresistivity changes dramatically with temperature and disorder. In order to explain such phenomenology, in this work, we extend our previous analyses on multiband transverse magnetoresistivity in magnesium diboride, by calculating its analytic expression, assuming a constant anisotropic Fermi surface mass tensor. The calculation is done for arbitrary orientation of the magnetic field with respect to the crystalline axes and for the current density either perpendicular or parallel to the magnetic field. This approach allows to extract quite univocally the values of the scattering times in the $\sigma$- and $\pi$- bands by fitting experimental data with a simple analytic expression.

We also extend the analysis to the magnetoresistivity of polycrystalline samples, with an arbitrary angle between the current density and the magnetic field, taking into account the anisotropy of each randomly oriented grain.

Thereby, we propose magnetoresistivity as a very powerful characterization tool to explore the effect of disorder by irradiation or selective doping as well as of phonon scattering in each one of the two types of bands, in single crystals and polycrystalline samples, which is a crucial issue in the study of magnesium diboride.

# 1. Introduction

The study of superconductivity in magnesium diboride has shown that the presence of two types of bands crossing the Fermi level has many implications in terms of normal state and superconducting properties[1,2]. Two bands, the $\pi 1$- and $\pi 2$-bands, are formed by the $p_z$ orbitals of boron atoms; they are weakly coupled to the phonons, they have three-dimensional character, one of them has electron-like charge carriers and the other one has hole-like ones. The other two bands, the $\sigma 1$- and $\sigma 2$-bands, are formed by $sp^2$-hybridized orbitals stretched along boron-boron bonds and are two-dimensional, hole-type and strongly coupled with the optical $E_{2g}$ phonon mode. Due to inhibition of interband scattering by the different parity of the $\pi$ and $\sigma$ types of bands, transport may occur predominantly in either $\pi$- or $\sigma$- bands depending on the ratio of the scattering times $\beta=\tau_\pi/\tau_\sigma$. This ratio is a crucial parameter in that many physical properties change dramatically depending on the $\pi$ or $\sigma$ character of the sample, for example, infrared reflectivity[3], microwave conductivity [4] and upper critical field[5,6]; hence the importance of having a simple experimental method of determining the ratio $\beta$.

Magnetoresistivity is a powerful tool for exploring directly the shape of the Fermi surface of metals, which is intimately related to transport coefficients as well as to equilibrium and optical properties. Moreover, the information about the electronic band structure contained in the Fermi surface, are of particular interest to theoreticians who carry out first-principles band structure calculations.

Magnesium diboride exhibits positive magnetoresistivity, which may be large in clean samples. Measured values are scattered within a broad range, depending on the sample purity and different scattering times ratios [7,8,9,10,11,12].

Due to its sensitivity to disorder, magnetoresistivity has proven to be a unique tool to determine the ratio of scattering times in $\pi$- and $\sigma$- bands in magnesium diboride polycrystalline [11] and epitaxial thin films [12]. In fact, it is not affected by uncertainties related to unknown geometrical factors as it is the case of resistivity in polycrystalline samples with poor intergrain connectivity, nor it requires interpretative models valid only within a limited range of parameters, as it is the case of upper critical field [13]. In particular, it has been shown that anisotropic magnetoresistivity measurements in epitaxial thin films exhibit a crossover: the magnetoresistivity in the configuration with **B** perpendicular to the *ab* planes is larger than in the configuration with **B** parallel to the *ab* planes for small $\beta$ values, while it becomes smaller with increasing $\beta$, that means as $\sigma$-bands become dirtier relative to $\pi$-bands.

Recent experimental data of magnetoresistivity as a function of magnetic field, magnetic field orientation and temperature, carried out on extremely clean epitaxial thin films[14], have shown a pretty rich phenomenology. At low temperatures just above the superconducting transition, the magnetoresistivity is minimum when the field is parallel to the *ab* planes, it reaches a maximum at intermediate angles and it slightly decreases again as the field direction approaches 90° with respect to the *ab* planes. At larger temperatures this trend reverses and eventually at 120K the opposite behavior is observed: the magnetoresistance is maximum when the field is parallel to the *ab* planes and minimum when it is perpendicular.

Motivated by these experimental findings, we calculate the anisotropic transverse magnetoresistivity, with the magnetic field forming a generic angle $\vartheta$ with boron planes (*ab* planes in the following). Using an analytical expression derived from the Boltzmann equation, assuming a constant anisotropic mass tensor on each Fermi surface sheet, with different in-plane and out-of-plane effective masses, we reproduce the rich phenomenology of experimental magnetoresistivity data.

Furthermore, we extend this approach to bulk samples with randomly oriented grains and we propose to carry out experimental measurements of both longitudinal and transverse magnetoresistivity, to which the predicted results should be applied.

## 2. Calculation of magnetoresistivity of single crystals

As it is well known [15], for a metal with a single, spherically symmetric band, the standard Boltzmann's equation approach yields no transverse magnetoresistivity. Oppositely, when two or more bands cross the Fermi level and the charge carriers carry out significant portions of cyclotron orbits before being scattered from one band to another, a non vanishing positive magnetoresistivity appears. In general, the contributions to magnetoresistivity of bands with different types of carriers - holes or electrons - add, whereas the contributions of bands with the same type of carriers subtract and eventually vanish, if the respective mobilities are identical. This makes the multiband magnetoresistivity dramatically sensitive to the presence of bands with different type of carriers or different mobilities. For this reason, in the case of magnesium diboride, we have necessarily to include the contribution of all of the four bands. In particular, the fact that $\pi 1$- and $\pi 2$-bands are electron-like and hole-like respectively, as well as the fact that $\sigma 1$- and $\sigma 2$-bands have different effective masses seriously affects magnetoresistivity, making a two-band calculation inadequate to describe experimental data: all the four bands have to be taken into account.

From Boltzmann equation, in presence of electric and magnetic fields **E** and **B**, whose intensity is not large enough that quantum effects should be considered, for each band $i$, the following equation for the unknown current density $\mathbf{J}_i$ is obtained [16]:

$$\mathbf{E} = \widetilde{\sigma}_i^{-1} \mathbf{J}_i + \mathbf{B} \times \widetilde{\mu}_i \widetilde{\sigma}_i^{-1} \mathbf{J}_i \qquad (1)$$

where $\widetilde{\sigma}_i^{-1}$ is the inverse conductivity tensor and $\widetilde{\mu}_i$ is the electric mobility tensor of the *i-th* band (in the following $\widetilde{\mu}_i$ has the units of an inverse magnetic field); they are related to each other by the carrier concentrations per unit cell in the *i-th* band $n_i$ by $\widetilde{\mu}_i = V\widetilde{\sigma}_i/(n_i e)$, being $V$ the unit cell volume and $e$ the electron charge with sign (positive for holes and negative for electrons). The three components of the current density can be obtained by summing the contributions of the corresponding components of all the bands. In order to extract the transverse magnetoresistivity in the configuration illustrated in figure 3(a) (configuration (1), with the current in the *ab* plane, along the x direction perpendicular to **B** ), the condition that the total current is zero both in the direction parallel to **B** and in the direction perpendicular to the **B**-x plane must be imposed: the resistivity $\rho(B)$ is obtained as the ratio of the electric field component along the direction of the applied current to the applied current density. The resulting expression for the total inverse total resistivity in configuration (1), is:

$$(\rho_{(ab)}(B))^{-1} = \sum_{i=\pi1,\pi2,\sigma1,\sigma2} \frac{\sigma_{i(ab)}}{g_i(B,\vartheta)} +$$

$$+ \frac{\left\{ B^2 \sin^2(\vartheta) \left( \sum_{i=\pi1,\pi2,\sigma1,\sigma2} \frac{\sigma_{i(ab)}\mu_{i(ab)}}{g_i(B,\vartheta)} \right) \left[ \left( \left( \sum_{i=\pi1,\pi2,\sigma1,\sigma2} \frac{\sigma_{i(c)}\left(1+\mu_{i(ab)}^2 \sin^2(\vartheta)B^2\right)}{g_i(B,\vartheta)} \right) \left( \sum_{i=\pi1,\pi2,\sigma1,\sigma2} \frac{\sigma_{i(ab)}\mu_{i(ab)}}{g_i(B,\vartheta)} \right) \right) + \left( B^2 \cos^2(\vartheta) \left( \sum_{i=\pi1,\pi2,\sigma1,\sigma2} \frac{\sigma_{i(c)}\mu_{i(ab)}}{g_i(B,\vartheta)} \right) \left( \sum_{i=\pi1,\pi2,\sigma1,\sigma2} \frac{\sigma_{i(c)}\mu_{i(ab)}^2}{g_i(B,\vartheta)} \right) \right) \right] \right. }{ \left( \left( \sum_{i=\pi1,\pi2,\sigma1,\sigma2} \frac{\sigma_{i(c)}\left(1+\mu_{i(ab)}^2 \sin^2(\vartheta)B^2\right)}{g_i(B,\vartheta)} \right) \left( \sum_{i=\pi1,\pi2,\sigma1,\sigma2} \frac{\sigma_{i(ab)}\left(1+\mu_{i(ab)}\mu_{i(c)} \cos^2(\vartheta)B^2\right)}{g_i(B,\vartheta)} \right) \right) - \left( B^4 \sin^2(\vartheta)\cos^2(\vartheta) \left( \sum_{i=\pi1,\pi2,\sigma1,\sigma2} \frac{\sigma_{i(c)}\mu_{i(ab)}^2}{g_i(B,\vartheta)} \right)^2 \right) }$$

$$\left. + B^2 \cos^2(\vartheta) \left( \sum_{i=\pi1,\pi2,\sigma1,\sigma2} \frac{\sigma_{i(c)}\mu_{i(ab)}}{g_i(B,\vartheta)} \right) \left[ \left( \left( \sum_{i=\pi1,\pi2,\sigma1,\sigma2} \frac{\sigma_{i(ab)}\left(1+\mu_{i(ab)}\mu_{i(c)}\cos^2(\vartheta)B^2\right)}{g_i(B,\vartheta)} \right) \left( \sum_{i=\pi1,\pi2,\sigma1,\sigma2} \frac{\sigma_{i(c)}\mu_{i(ab)}}{g_i(B,\vartheta)} \right) \right) + \left( B^2 \sin^2(\vartheta) \left( \sum_{i=\pi1,\pi2,\sigma1,\sigma2} \frac{\sigma_{i(ab)}\mu_{i(ab)}}{g_i(B,\vartheta)} \right) \left( \sum_{i=\pi1,\pi2,\sigma1,\sigma2} \frac{\sigma_{i(c)}\mu_{i(ab)}^2}{g_i(B,\vartheta)} \right) \right) \right] \right\}$$

where

$$g_i(B,\vartheta) = \left(1 + \mu_{i(c)}\mu_{i(ab)}B^2 \cos^2(\vartheta) + \mu_{i(ab)}^2 B^2 \sin^2(\vartheta)\right)$$

(2)

The first term in equation (2) is associated to the electric field parallel to the applied current ($E_x$ with reference to the axes of figure 3(a)), while the other terms are associated to the electric field components along the other transverse direction and along the direction of **B.** Approximate Fermi surface average values of the effective masses and carrier concentrations per unit cells are taken from band structure calculations of Profeta *et al.* [17]. The effective masses are defined for each band and cartesian direction as $\frac{1}{m_{i(\alpha)}} = \frac{m_0}{\hbar^2} \left\langle \frac{\partial^2 \varepsilon_i(\vec{k})}{\partial k_\alpha^2} \right\rangle_{\varepsilon_F}$ ($i = \pi1, \pi2, \sigma1, \sigma2; \alpha = x, y, z$), where

$\frac{1}{\hbar^2}\left\langle \frac{\partial^2 \varepsilon_i(\vec{k})}{\partial k_\alpha^2} \right\rangle_{\varepsilon_F} = \frac{V}{N_i(\varepsilon_F)} \int_{BZ} \frac{d\vec{k}}{4\pi^3} \frac{1}{\hbar^2} \frac{\partial^2 \varepsilon_i(\vec{k})}{\partial k_\alpha^2} \delta(\varepsilon_F - \varepsilon_i(\vec{k}))$ denotes the average value of the inverse mass tensor taken over the Fermi surface, $V$ is the unit cell volume and $N_i(\varepsilon)$ are the partial density of states per unit cell of each band $N_i(\varepsilon) = V \int_{BZ} \frac{d\vec{k}}{4\pi^3} \delta(\varepsilon - \varepsilon_i(\vec{k}))$ (the integral over $\vec{k}$ here is performed over the entire Brillouin zone (BZ)). The carrier concentrations per unit cell are $n_i = \int_{-\infty}^{\varepsilon_F} N_i(\varepsilon)d\varepsilon$ for electrons bands and $n_i = \int_{\varepsilon_F}^{\infty} N_i(\varepsilon)d\varepsilon$ for holes bands. The energy bands are interpolated by a spline fit [18] on the *ab initio* energy bands and the integration over the Fermi surface are performed by the linear tetrahedron method [19]. As a result, we assume for the effective masses and carrier densities the values listed in table I, where all the effective masses are expressed in units of the free electron mass $m_0$=9.1·10$^{-31}$ Kg. We note that the values of the effective masses in $\sigma$-bands along the $c$ axis may be questionable, indeed, the curvature in the tubular sheets even changes sign along the $c$ direction; actually, it is questionable the use of the effective mass approximation at all in this case. However, since such $m_{\sigma1(c)}$ and $m_{\sigma2(c)}$ values are much larger than all the other effective mass values, the contribution to transport of $\sigma$-bands along the $c$ direction is very small and the calculated numerical values of magnetoresistivity curves are largely independent of this choice as well.

In the general case, the matricial equation (1) can only be solved by expanding in series of $B$; instead, within our approximation of constant average mass tensor, equation (2) is obtained without any truncations, so that it is valid to all orders in $B$. Indeed, a saturation regime at high fields is reached, as presented in the following section.

Considering an average scattering time $\tau_\pi$ in $\pi_1$- and $\pi_2$-bands and similarly an average scattering time $\tau_\sigma$ in both $\sigma$-bands, all the mobility values scale as the inverse ratio of the respective masses multiplied by the ratio of the respective scattering times. The mobilities of the four bands can be expressed in terms of only two free parameters which we define as the ratio of the scattering times $\beta = \tau_\pi / \tau_\sigma$ and the scaling factor $\alpha = |\mu_{(c)\pi1}|$:

$$\mu_{i(dir)} = \pm \frac{\alpha \cdot m_{\pi 1(c)}}{m_{i(dir)}} \frac{\tau_i}{\tau_\pi} \qquad \text{where } i = \pi 1, \pi 2, \sigma 1 \text{ or } \sigma 2 \text{ and } (dir) = (ab) \text{ or } (c) \qquad (3)$$

The charge carrier densities $n_i$ enter the calculation by relating the mobility and the conductivity of each band and direction $\mu_{i(dir)} = V\sigma_{i(dir)}/(n_i e)$ (where $i=\pi 1, \pi 2, \sigma 1$ or $\sigma 2$ and $(dir)=(ab)$ or $(c)$).

The fitting parameters α and β determine not only magnetoresistivity curves, but also the calculated total in-plane resistivity of the sample in zero field $\rho_{(ab)}$:

$$\rho_{(ab)} = \frac{e}{V} \left( \sum_{i=\pi 1, \pi 2, \sigma 1, \sigma 2} |\mu_{i(ab)} n_i| \right)^{-1} \qquad (4)$$

Its comparison with the experimental resistivity value in zero field provides a further constraint in the fitting.

**3. Results and discussion**

In figure 1, the curves for the magnetoresistivity $(\rho(B) - \rho(0))/\rho(0)$ obtained from equation (2) are plotted as a function of the squared magnetic field for three different angles between the applied field **B** and the *ab* planes and for different values of the parameters $\beta=\tau_\pi/\tau_\sigma$ and $\alpha=|\mu_{(c)\pi 1}|$. In particular β increases from top to bottom, that is the σ-bands become increasingly dirty with respect to the π ones; the parameter α is chosen in such a way that the in-plane resistivity of the sample $\rho_{(ab)}$ given by equation (4) is 1 μΩ·cm in all cases, which is a typical value for clean thin films [20]. It can be seen that for low values of β the magnetoresistivity in the configuration **B**⊥*ab* (ϑ=90°) is larger than in the configuration **B**//*ab* (ϑ=0°) while for large enough β values the opposite is true. Moreover, for intermediate values of β the magnetoresistivity is non monotonic with the angle ϑ. Finally, it is worth to note that in the configuration **B**//*ab* (ϑ=0°), the saturation of magnetoresistivity is already evident, especially for large β values, while at the same fields in the **B**⊥*ab* (ϑ=90°) configuration the magnetoresistivity is still almost linear with *B* squared.

In equation (2) the quantity $g_i(B,\vartheta)$ is related to the cyclotron frequency of the $i$-th band in the plane perpendicular to the magnetic field $\omega_i$:

$$g_i(B,\vartheta) = \left(1 + \mu_{i(c)}\mu_{i(ab)}B^2\cos^2(\vartheta) + \mu_{i(ab)}^2 B^2\sin^2(\vartheta)\right) = 1 + \omega_i^2\tau_i^2$$

$$\omega_i(B,\vartheta) = \frac{eB}{m_{i(ab)}}\sqrt{\sin^2(\vartheta) + \frac{m_{i(ab)}}{m_{i(c)}}\cos^2(\vartheta)}$$

(5)

Indeed, for each $i$-th band, the derivative of the area of the Fermi surface cross-section with respect to the energy, $\partial A^{(i)}/\partial\varepsilon$, is proportional to the cyclotron mass $m_{cycl}^{(i)}$ in the same plane. In turns, the cyclotron mass is inversely related to the cyclotron frequency $\omega_i$, so that:

$$\omega_i = \left(2\pi eB/\hbar^2\right)\left(\partial A^{(i)}/\partial\varepsilon\right)^{-1}$$

(6)

The derivatives $\partial A^{(i)}/\partial\varepsilon$ can be easily calculated in the assumed approximation of ellipsoidal Fermi surface, giving the expression (5) for $\omega_i(B,\vartheta)$. When $\mathbf{B}\perp ab$ the only relevant cyclotron frequencies are the in-plane ones $\omega_{i\perp} = \frac{eB}{m_{cycl\perp}^{(i)}} = \frac{eB}{m_{i(ab)}}$, whereas in the case $\mathbf{B}//ab$ the magnetoresistivity is mainly determined by cyclotron frequencies $\omega_{i\|} = \frac{eB}{m_{cycl\|}^{(i)}} = \frac{eB}{\sqrt{m_{i(c)}m_{i(ab)}}}$. For a very clean sample with comparable scattering times in all the bands at low temperature ($\beta$ close to unity), the relative magnitudes of the cyclotron frequencies in the different bands determine the angular dependence of magnetoresistivity. When $\mathbf{B}\perp ab$, the cyclotron frequency $\omega_{i\perp} \propto \frac{1}{m_{i(ab)}}$ is larger for the $\sigma$-bands than for the $\pi$-bands, as shown by the effective mass values listed in table I. This can be understood physically, since the $\sigma$-bands carriers orbit the small circumference of the tubular Fermi surfaces, while the $\pi$– bands carriers orbit the large circle of the honeycomb around the $\Gamma$ point. The opposite is true for $\mathbf{B}//ab$, where the $\sigma$-band orbits are open (our cyclotron mass is indeed very large and thereby the cyclotron frequency $\omega_{i//}$ of $\sigma$-electron is very small) and the $\pi$ carriers orbit the small circumference of the tube forming the honeycomb Fermi surfaces (larger cyclotron frequency $\omega_{i//}$ of $\pi$-electrons). As a result, the magnetoresistivity is maximum in the $\mathbf{B}\perp ab$ configuration and it decreases with decreasing angle $\vartheta$. The above argument is still true for samples with cleaner $\sigma$-bands ($\beta$ smaller than unity), but it reverses when transport occurs predominantly in $\pi$-bands ($\beta$ significantly larger than unity). In the latter limit, the magnetoresistivity is maximum for $\vartheta=0°$ and it decreases with increasing angle $\vartheta$, due to the larger cyclotron masses $m_{cycl\perp}^{(\pi 1)}$ and $m_{cycl\perp}^{(\pi 2)}$ with respect to $m_{cycl//}^{(\pi 1)}$ and $m_{cycl//}^{(\pi 2)}$ (see the list of effective masses given in table I, keeping in mind that $m_{cycl//}^{(i)} = \sqrt{m_{i(c)} m_{i(ab)}}$, while $m_{cycl\perp}^{(i)} = m_{i(ab)}$). This argument on the relative magnitudes of these cyclotron masses explains also the reason why, at large $\beta$, the saturation regime is reached at smaller fields in the $\boldsymbol{B}//ab$ ($\vartheta=0°$) configuration than in the $\boldsymbol{B}\perp ab$ ($\vartheta=90°$) configuration, being $\omega_{i//}$ larger than $\omega_{i\perp}$ for $\pi$-bands.

The crossover between the magnetoresistivity in the $\boldsymbol{B}//ab$ and $\boldsymbol{B}\perp ab$ configurations as a function of $\beta$ has been indeed experimentally observed by measuring magnetoresistivity at 42K in magnetic fields up to 45 Tesla in films with different amounts of disorder introduced by neutron irradiation or carbon doping [12]. The crossover value of $\beta$ between the limiting cases of increasing and decreasing $\vartheta$-dependence of magnetoresistivity is somewhat larger than unity, as it can be seen from figure 1, in the case of samples with resistivity $\rho_{(ab)}=1$ $\mu\Omega\cdot$cm; this is due to the fact that in-plane $\sigma$ effective masses are smaller than in-plane $\pi$ effective masses, so that the product of cyclotron frequencies and scattering times is nearly the same for the two types of bands when $\beta$ is slightly larger than unity.

As a consequence of the presence of open orbits in the Fermi surface, it could happen that the experimental magnetoresistivity of clean samples does not saturate even at very large fields. Of course this effect could not be accounted for in our approximation of constant average mass tensor. However the fair agreement between the results of this analytical model and experimental measurements [12,14] indicates that neglecting the contribution of open orbits is acceptable.

It is very important to check the sensitivity of our approach to uncertainties in the effective masses relative values, keeping in mind that only the relative ratios between them and not their absolute values enter the equations. For example, a direct test shows that a 10% change in the four σ effective masses with respect to the π ones yields a variation in the calculated magnetoresistivity of 3% for B parallel to the *ab* planes and 11% for B along the *c* axis, whereas a 10% change in the four π effective masses with respect to the σ ones yields a variation of the calculated magnetoresistivity of 5% for B parallel to the *ab* planes and 10% for B along the *c* axis. The effective mass values obtained from band structure calculations may indeed suffer from systematic uncertainties, but it is not easy to set a realistic upper limit to such error. Unfortunately, experimental data of all the eight effective masses of the four bands along *ab* planes and *c* axis are not available. Measured values of $m_{\sigma1(ab)}$, $m_{\sigma2(ab)}$, $m_{cycl//}^{(\pi1)}$ and $m_{cycl//}^{(\pi2)}$ extracted from de Haas-van Alphen experiments [21] correspond to $m_{\sigma1(ab)}/m_{\sigma2(ab)}$ and $m_{cycl//}^{(\pi1)}/m_{cycl//}^{(\pi2)}$ ratios which agree within few percent with our calculated ratios; instead, the ratios between measured σ effective masses and π cyclotron masses differ significantly from the calculated ones. This discrepancy is not surprising if we consider that the calculated masses are average values of the inverse mass tensors taken over the Fermi surface while in the case of de Haas-van Alphen experiments the masses are those of the extremal orbits only. In the case of σ bands for *B* perpendicular to the *ab* planes, the extremal orbits are representative of the whole Fermi surface, but the same thing cannot be said of the tubular network shape of π bands.

The curves shown in figure 1 can be used to fit experimental magnetoresistivity and resistivity data, thus extracting quite univocally the values of the scattering times in the σ- and π- bands. This method can be used to study the effect of disorder, irradiation and selective doping in the two bands, which is a crucial issue in the research on magnesium diboride. Also, transport as a function of temperature can be studied: the measurement of magnetoresistivity curves with increasing temperature allows to extract the values of the scattering rates in σ– and π– bands in different temperature regimes where either impurity scattering or phonon scattering dominate. For example, it is expected that a clean sample with comparable scattering times in the two types of bands at low temperature will exhibit the crossover between larger magnetoresistivity for **B**⊥*ab* or **B**//*ab* with increasing temperature, due to the much stronger coupling of $\sigma$ carriers with phonons, which switches the conduction from the $\sigma$ to the $\pi$ channel with increasing temperature. Such measurements require highly oriented epitaxial films. Moreover, unless the films are extremely clean, large magnetic fields must be reached to observe a significant magnetoresistivity.

For a closer inspection of the angular dependence of magnetoresistivity, in figure 2 we plot magnetoresistivity as a function of $\vartheta$ for different values of magnetic fields, namely *B*=9T, *B*=20T and *B*=45T, which are typical values of field easily reached by commercial superconducting magnets, copper magnets and pulsed magnets, respectively. Again, the parameter $\beta$ increases from the top to the bottom panel, that is with the $\sigma$-bands becoming increasingly dirty with respect to the $\pi$ ones. The curves reproduce very well the experimental behaviour [14]: at low $\beta$ (cleaner $\sigma$-bands) the magnetoresistivity increases with $\vartheta$, at large enough $\beta$ (cleaner $\pi$-bands) it decreases with $\vartheta$ and at intermediate $\beta$ close to unity it shows a non monotonic behaviour, with a maximum whose angular position depends not only on the parameters $\beta$ and $\alpha$, but also on the intensity of the magnetic field. Indeed, the results at different fields are qualitatively very similar, except for the magnetoresistivity absolute value which is obviously larger at larger fields and for the position of the maximum, whose shift to smaller angles with increasing β is faster at small fields and slowest at larger fields, where the saturation regime in the ***B***//*ab* ($\vartheta=0°$) configuration is well reached.

We note that as a consequence of the anisotropic effective mass, even in the case that the applied current is *parallel* to the magnetic field there would exist a non zero magnetoresistivity ($\rho_{//B}(B) - \rho_{//B}(B=0))/\rho_{//B}(B=0)$) (see configuration (3) in figure 3(c)) [22]. This is due to the tensorial relationship between the current density and the electric field, which couples all the three spatial components of equation (1). In figure 3(d), this longitudinal magnetoresistivity is plotted as a function of the angle $\vartheta$ between the *ab* planes and the direction parallel to both the magnetic field and the applied current, for *B*=45T and for three values of the parameter $\beta$ (also in this case the corresponding values of the parameter $\alpha=|\mu_{\pi 1}^{(c)}|$ are fixed in such a way that the in-plane resistivity turns out to be $\rho_{(ab)}=1$ μΩ·cm in all cases). It can be seen that it vanishes in the limiting cases $\vartheta=0°$ and $\vartheta=90°$, but it assumes finite values at intermediate angles, being larger for small values of $\beta$. This is indeed expected, because this is an effect related to the anisotropic conductivity and it is enhanced when transport occurs mainly in the σ-bands (low $\beta$) which are the most anisotropic. For comparison, the longitudinal magnetoresistivity is plotted together with the above described transverse magnetoresistivity (configuration (1) in figure 3(a)), for the same values of $B$ and $\beta$. It can be seen that the longitudinal contribution is not negligible, especially for small β; in particular, for $\vartheta\sim30°$, $\beta=2$ and $B=45$T the longitudinal magnetoresistivity is only 2.5 times smaller than the transverse magnetoresistivity in configuration (1). Instead, for larger $\beta$ and/or for $\vartheta$ approaching 0° or 90° the longitudinal magnetoresistivity becomes vanishing small. Even if it is apparent that it is impossible in practice to realise an experimental configuration with the applied magnetic field and current forming an angle $\vartheta$ with $ab$ crystalline planes in a single crystal or in a $c$-oriented $MgB_2$ film, there exists the possibility of setting up such configuration in epitaxial films whose $c$ axis is not parallel to the growth direction, such as in the case of films grown on Yttrium Stabilised Zirconia substrates, whose $c$-axis is tilted by 58 degrees with respect to the substrate plane [23].

Besides the above described longitudinal and transverse configurations, there exists a third possible transverse configuration, indicated as configuration (2) in figure 3(b), where the applied current is perpendicular to **B** and forms an angle ($\vartheta-\pi/2$) with the $ab$ planes. As seen in figure 3(d), the transverse magnetoresistivity in this configuration (2) merges with the one in configuration (1) when $\vartheta$ approaches 90° as expected, because when **B** is perpendicular to the $ab$ planes the system has cylindrical symmetry and the two configurations are indistinguishable. At smaller $\vartheta$, instead, the current direction is nearly parallel to the $c$-axis; if transport occurs mainly in σ-bands (small $\beta$), the magnetoresistivity is smaller in the transverse configuration (2), due to the smaller zero field conductivity along the $c$ axis in σ-bands, while if transport occurs mainly in π-bands (large $\beta$), it is smaller in the transverse configuration (1), due to the larger zero field conductivity along the *c*-axis in *π*-bands.

**4. Magnetoresistivity of polycrystalline samples**

Let us consider the case of polycrystalline samples. In polycrystals, the *ab* planes of each grain are randomly oriented with respect to the magnetic field and the local percolative current density forms an arbitrary angle with the magnetic field. A quantitative analysis of transverse magnetoresistivity in polycrystals has been carried out within an average isotropic effective mass approximation [11]. This isotropic approach yields zero longitudinal magnetoresistivity. Here, we go beyond this approximation and we show that in certain cases the isotropic approach is inadequate in describing the magnetoresistivity behaviour. Indeed, non vanishing longitudinal magnetoresistivity has been observed in polycrystalline metals with anisotropic Fermi surface shape [24].

Starting from equation (1), we assume again an anisotropic mass tensor and a *homogeneous electric field*, that is a local electric field parallel to the average electric field. The latter simplification is valid only in the case of extremely dense samples with high intergrain connectivity. In the opposite limit, the local electric field is determined for each grain also by the shape of the grain itself as well as by the number of adjacent neighbouring grains, which makes the problem very complicated to treat. We consider an arbitrary angle $\vartheta$ between **B** and the *ab* planes and we calculate the current density **J** along any arbitrary direction. After averaging over $\vartheta$, the transverse magnetoresistivity is obtained by imposing that the average **J** is zero along the direction parallel to **B**, while the longitudinal magnetoresistivity is obtained by imposing that the average **J** is zero in the plane perpendicular to **B**. In the general case the expression for the inverse resistivity is:

$$\rho(B,\delta)^{-1} = \frac{R\left[P^2+Q^2\right]}{\sin^2(\delta)\,P\,R + \cos^2(\delta)\left[P^2+Q^2\right]}$$

where

$$P(B) = \frac{1}{4} \int_{-\pi/2}^{\pi/2} \left( \sum_{i=\pi1,\pi2,\sigma1,\sigma2} \frac{\sigma_{i(c)}\cos^2(\vartheta) + \sigma_{i(ab)}\sin^2(\vartheta) + \sigma_{i(ab)}}{1 + \mu_{i(c)}\mu_{i(ab)}B^2\cos^2(\vartheta) + \mu_{i(ab)}^2 B^2\sin^2(\vartheta)} \right) \cos(\vartheta) d\vartheta$$

$$Q(B) = \frac{1}{2} \int_{-\pi/2}^{\pi/2} \left( \sum_{i=\pi1,\pi2,\sigma1,\sigma2} \mu_{i(ab)} B \frac{\sigma_{i(c)}\cos^2(\vartheta) + \sigma_{i(ab)}\sin^2(\vartheta)}{1 + \mu_{i(c)}\mu_{i(ab)}B^2\cos^2(\vartheta) + \mu_{i(ab)}^2 B^2\sin^2(\vartheta)} \right) \cos(\vartheta) d\vartheta \qquad (7)$$

$$R(B) = \frac{1}{2} \int_{-\pi/2}^{\pi/2} \left( \sum_{i=\pi1,\pi2,\sigma1,\sigma2} \frac{\sigma_{i(c)}\sin^2(\vartheta) + \sigma_{i(ab)}\cos^2(\vartheta) + \sigma_{i(c)}\mu_{i(ab)}^2 B^2}{1 + \mu_{i(c)}\mu_{i(ab)}B^2\cos^2(\vartheta) + \mu_{i(ab)}^2 B^2\sin^2(\vartheta)} \right) \cos(\vartheta) d\vartheta$$

Here $\delta$ is the angle between the magnetic field **B** and the macroscopic applied current **J** ($\delta=0°$ for longitudinal magnetoresistivity and $\delta=90°$ for transverse magnetoresistivity). In figure 4(a) the longitudinal and transverse magnetoresistivities of bulk samples are plotted as continuous lines for three values of the parameter $\beta$ (again, the values of the parameter $\alpha=|\mu_{\pi1}^{(c)}|$ are fixed in such a way that the zero field resistivity turns out to be $1\ \mu\Omega\cdot$cm in all cases). The transverse magnetoresistivity (**J**⊥**B**) slightly increases with increasing $\beta$, because at low $\beta$ transport is dominated by $\sigma$-bands and the low mobility of $\sigma$ carriers along the $c$ direction suppresses the average magnetoresistivity. On the contrary, the longitudinal magnetoresistivity (**J**∥**B**) tends to vanish at large $\beta$, because when transport is dominated by $\pi$-bands the system is close to be isotropic, while a significant longitudinal magnetoresistivity exists for small $\beta$. As a consequence of the above described argument, the magnetoresistivity dependence on the angle $\delta$ between the magnetic field and the applied current density is steeper for large $\beta$ and smoother for low $\beta$, as shown in figure 4(b).

Oppositely, in the isotropic approximation, assuming for each $i$-th band a diagonal mass tensor whose inverse elements are defined as:

$$1/m_i = \frac{2/m_{i(ab)} + 1/m_{i(c)}}{3} \qquad i = \pi1, \pi2, \sigma1 \text{ or } \sigma2 \qquad (8)$$

the isotropic inverse resistivity as a function of the angle $\delta$ between **J** and **B** is expressed as:

$$\rho(B,\delta)^{-1} = \frac{V[S^2 + T^2]}{\sin^2(\delta) SV + \cos^2(\delta)[S^2 + T^2]}$$

where

$$S(B) = \sum_{i=\pi 1,\pi 2,\sigma 1,\sigma 2} \frac{\sigma_i}{1+\mu_i^2 B^2}$$

$$T(B) = \sum_{i=\pi 1,\pi 2,\sigma 1,\sigma 2} \frac{\sigma_i \mu_i B}{1+\mu_i^2 B^2} \qquad (9)$$

$$V = \sum_{i=\pi 1,\pi 2,\sigma 1,\sigma 2} \sigma_i$$

It is clearly seen that the resistivity in the direction of the magnetic field ($\delta=0°$) does not depend on $B$ itself, yielding zero longitudinal magnetoresistivity. For comparison, in figures 4(a) and 4(b) the dotted curves calculated within the isotropic approximation are shown. In figure 4(a) it can be seen that the isotropic approximation tends to overestimate the transverse magnetoresistivity in all cases and for $\beta$ values larger than *1.5* the isotropic magnetoresistivity depends very weakly on $\beta$. Moreover the isotropic approach predicts zero longitudinal magnetoresistivity, while within the anisotropic averaged approach the longitudinal magnetoresistivity may turn out to be comparable to the transverse magnetoresistivity, especially for low $\beta$; for example, in the uppermost panel of figure 4(a) the ratio of the two magnetoresistivities is nearly *2.3*. Consistently, in figure 4(b) the angular dependence of the isotropic magnetoresistivity curves (dotted curves) is steeper, because the isotropic approximation predicts zero longitudinal contribution and it overestimates the transverse contribution.

This result is particularly noteworthy, in that the combined measurement of transverse and longitudinal magnetoresistivity in bulk samples provides enough input to extract unambiguously the β parameter, even if the sample resistivity is not known, due to the uncertainty on the geometrical factor. Indeed, to the best of our knowledge, no combined measurement of transverse and longitudinal magnetoresistivity in bulk samples have ever been presented in literature; we propose to carry out such experiment in order to characterise the scattering rates in different bands in $MgB_2$ bulk samples. This requires of course either very clean samples or very large magnetic fields. In addition, in order to apply equations (7), the samples should have high density and intergrain connectivity as well; otherwise, the assumption of homogeneous electric field fails, as long as the local electric field is not parallel to the average one, being determined by the grain shape, the connected adjacent grains and the conductivity mismatch at the interface between differently oriented adjacent grains.

**5. Conclusions**

We demonstrate that experimental findings on magnesium diboride normal state magnetoresistivity can be well accounted for within a cyclotron orbits scenario, in the free electron approximation, assuming a constant average mass tensor with different in-plane and out-of-plane components. Starting from the Boltzmann equation, we extract an analytic expression for the transverse magnetoresistivity, which allows to extract easily and univocally the scattering times in the $\sigma$– and $\pi$–bands from experimental data. We discuss the angular dependence of transverse and longitudinal magnetoresistivity in terms of the shape of the four-sheet Fermi surface of magnesium diboride. We extend this approach to polycrystalline samples by suitably averaging the anisotropy of each randomly oriented grain and we extract simple expressions for transverse and longitudinal magnetoresistivity.

Finally, we propose magnetoresistivity measurements with the current either parallel or perpendicular to the magnetic field as a univocal method to extract the scattering times in the two types of bands of magnesium diboride films, single crystal and polycrystalline samples. The magnetoresistivity can be measured at different temperatures, allowing to extract the temperature dependence of the scattering rates as well.

**Figure captions**

**Figure 1:** Calculated magnetoresistivity curves for the **B**⊥*ab* ($\vartheta=90°$), *B*//*ab* ($\vartheta=0°$) and $\vartheta=45°$ configurations for 4 different values of the parameter $\beta=\tau_\pi/\tau_\sigma$ equal to *0.2, 1.5, 5.0* and *10.0* from the top panel to the bottom one and for values of the parameter $\alpha=|\mu_{\pi 1}^{(c)}|$ equal to *0.0157, 0.0480, 0.0621* and *0.0661* m$^2\cdot$V$^{-1}\cdot$s$^{-1}$ respectively, such that the in-plane resistivity given by equation (4) turns out to be $\rho_{(ab)}=1$ μΩ·cm in all cases. In other words, from top to bottom the $\sigma$-bands become increasingly dirty with respect to $\pi$-bands.

**Figure 2:** Angular dependence of magnetoresistivity for *B*=9T, 20T and 45T and for the same parameters of figure 1, that is $\beta=\tau_\pi/\tau_\sigma$ equal to *0.2, 1.5, 5.0* and *10.0* from the top panel to the bottom one and $\alpha=|\mu_{\pi 1}^{(c)}|$ equal to *0.0157, 0.0480, 0.0621* and *0.0661* m$^2\cdot$V$^{-1}\cdot$s$^{-1}$ respectively, such that the in-plane resistivity given by equation (4) turns out to be $\rho_{(ab)}=1$ μΩ·cm in all cases.

**Figure 3:** Schematic sketches of the magnetic field and current orientation with respect to the crystalline axes in the case of (a) transverse magnetoresistivity with the current density in the *ab* planes, (b) transverse magnetoresistivity with the current density in the plane of **B** and the *c*-axis, (c) longitudinal magnetoresistivity. In the (d) panel, the magnetoresistivity in the three configurations is plotted as a function of the angle $\vartheta$ between the magnetic field and the *ab* planes for *B*=45T and for 3 different values of the parameter $\beta=\tau_\pi/\tau_\sigma$ equal to *0.2, 1.5,* and *5.0* and for values of the parameter $\alpha=|\mu_{\pi 1}^{(c)}|$ equal to *0.0157, 0.0480,* and *0.0621* m$^2\cdot$V$^{-1}\cdot$s$^{-1}$ respectively, such that the in-plane resistivity turns out to be $\rho_{(ab)}=1$ μΩ·cm in all cases. The curves corresponding to the transverse configuration (1) are portions of the plots of figure 2, for *B*=45T, in the range of $\vartheta$ between 0 and 90°.

**Figure 4:** Upper panel: transverse ($J \perp B$, $\delta=90°$) and longitudinal ($J \| B$, $\delta=0°$) magnetoresistivity curves of bulk samples for 3 different values of the parameter $\beta = \tau_\pi / \tau_\sigma$ equal to *0.2*, *1.5*, and *5.0* and for values of the parameter $\alpha = |\mu_{\pi1}^{(c)}|$ equal to *0.0193*, *0.0434*, and *0.0501* m$^2 \cdot$V$^{-1} \cdot$s$^{-1}$ respectively, such that the zero field bulk resistivity turns out to be *1* μΩ·cm in all cases; the curves are calculated either by taking into account the anisotropy of the randomly oriented grains (continuous lines) or within a isotropic approximation (open symbols). Lower panel: magnetoresistivity of bulk samples at *B*=45T and for the same parameters as above as a function of the angle $\delta$ between the magnetic field and the applied current density, calculated either by taking into account the anisotropy of the randomly oriented grains (continuous lines) or within a isotropic approximation (open symbols).

**Table I:** Effective masses for both crystal directions, as well as charge densities per unit cell, for all four bands of MgB$_2$ crossing the Fermi surface.

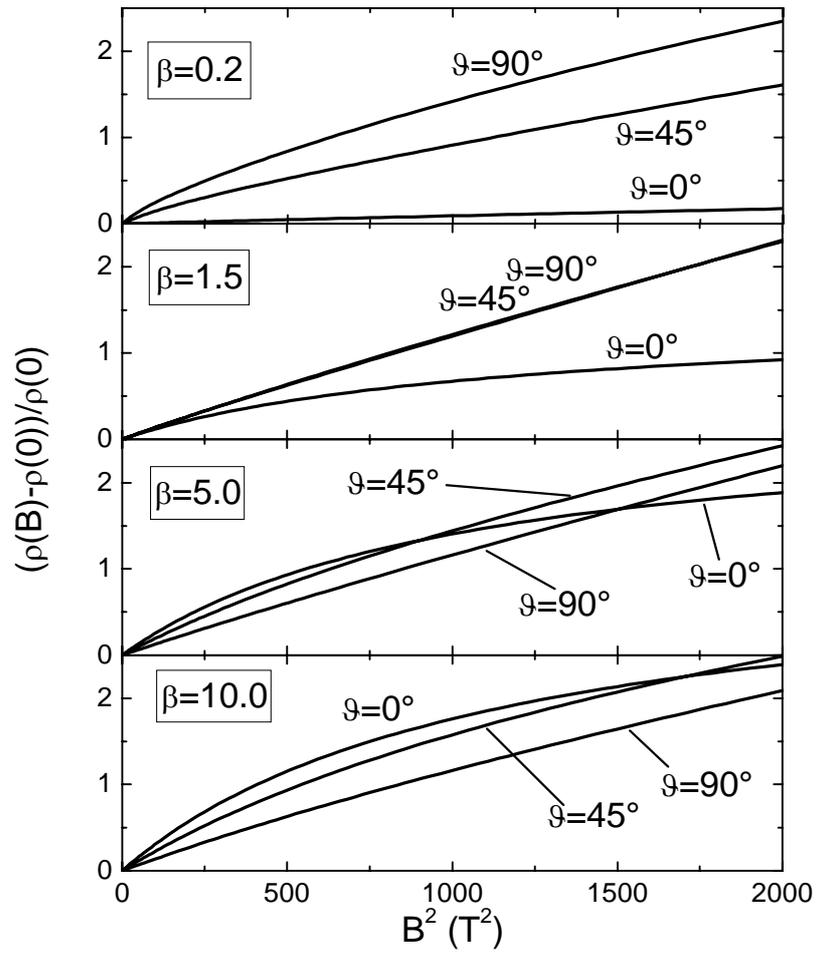

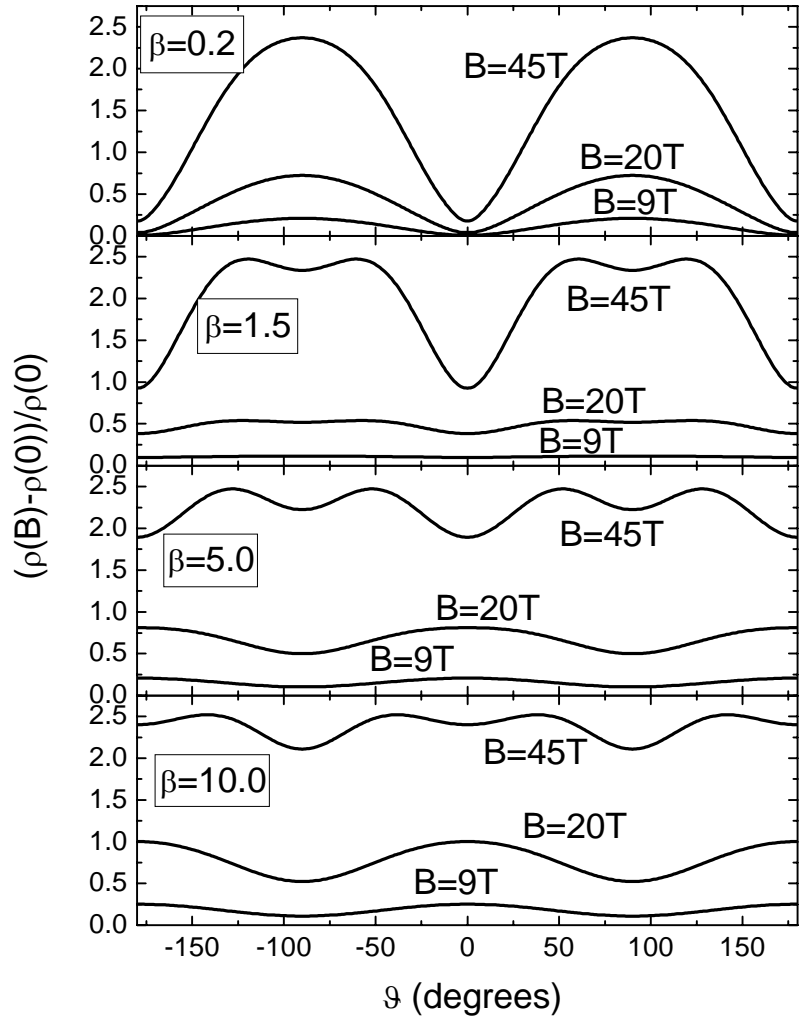

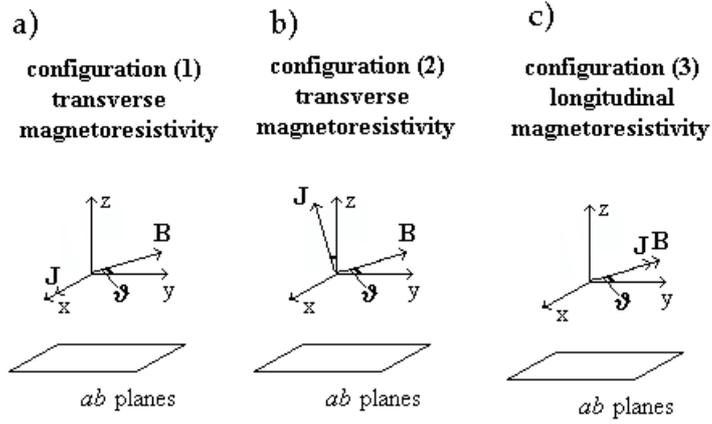
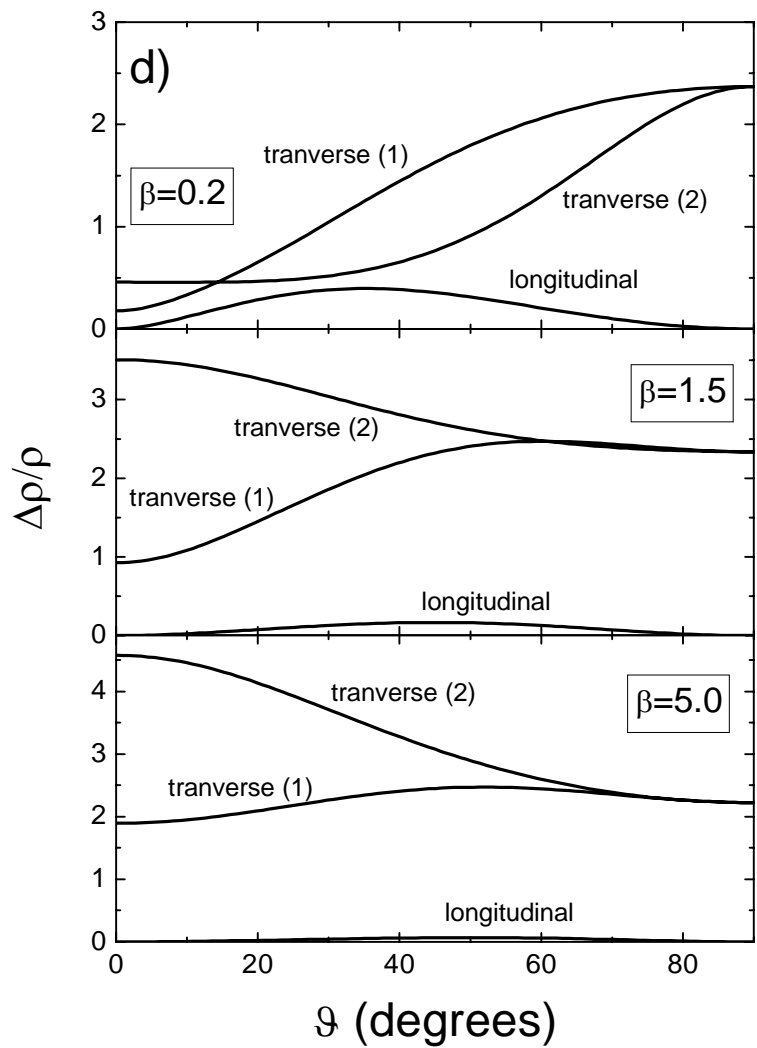

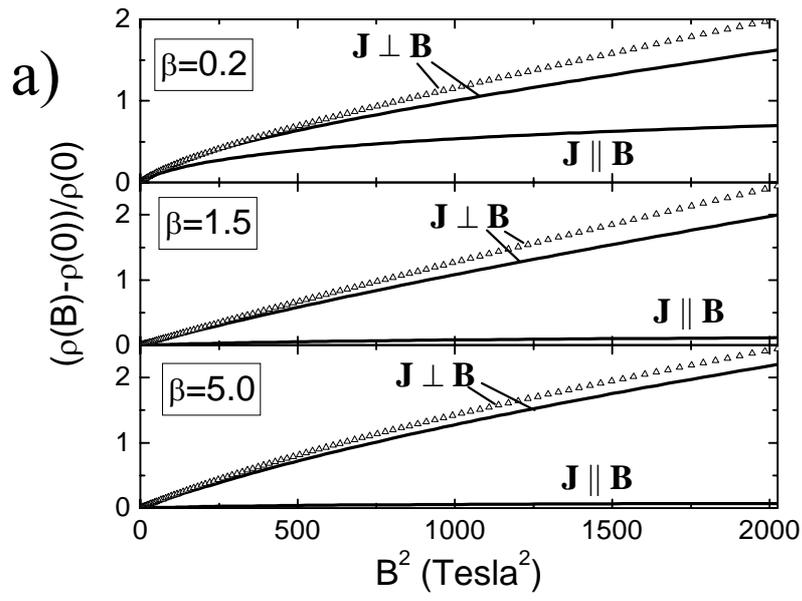

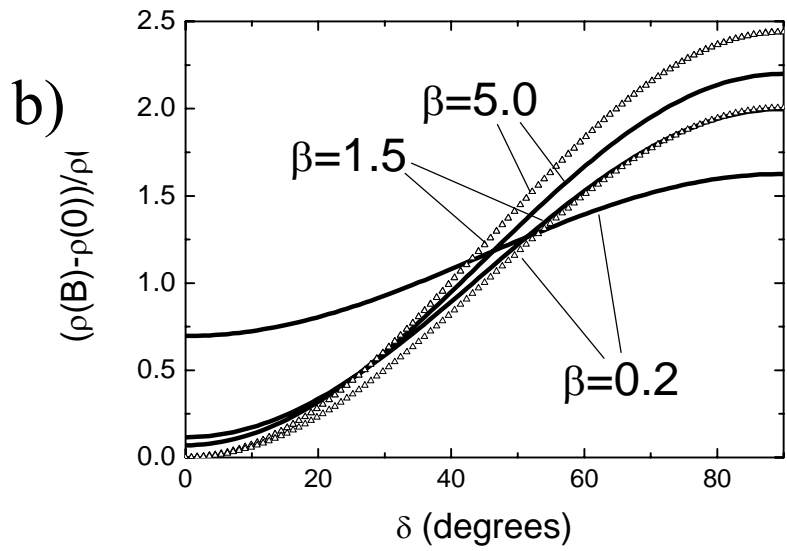

**Table I**

| Band | $m_{eff}/m_0$ | | n |
|---|---|---|---|
| | // ab planes | // c axis | |
| σ1 | 0.335 | 66.41 | 0.051 |
| σ2 | 0.755 | 212.02 | 0.100 |
| π1 | 0.879 | 0.628 | 0.265 |
| π2 | 1.077 | 0.305 | 0.114 |